\documentclass[aps,prd,preprint,groupedaddress]{revtex4-1}

\usepackage{axodraw}
\usepackage{epsfig}
\usepackage{xspace}
\usepackage{graphics}
\usepackage[utf8]{inputenc}
\usepackage{ifthen}
\usepackage{float}
\usepackage{multirow}

\newcommand{\pythia}{\textsc{Pythia}\xspace}
\newcommand{\dipsy}{\protect\scalebox{0.8}{DIPSY}\xspace}

\providecommand{\eqref}[1]{eq.~(\ref{#1})\xspace}

\renewcommand{\eqref}[1]{eq.~(\ref{#1})\xspace}

\newcommand{\fig}[1]{\ref{#1}}
\newcommand{\figref}[1]{figure~\fig{#1}}

\def\eg{\emph{e.g.} }

\begin{document} 

\title{Effects of Colour Reconnection on Hadron Flavour Observables
  \thanks{
    Work supported in part by the MCnetITN FP7 Marie Curie Initial
    Training Network, contract PITN-GA-2012-315877 and the Swedish Research
    Council, contract 621-2013-4287.
    	}
    }

%% %simple case: 2 authors, same institution
\author{Christian Bierlich and}
\author{Jesper Roy Christiansen}

\affiliation{Dept.~of Astronomy and Theoretical Physics, Lund University, Sweden}

\email{christian.bierlich@thep.lu.se}
\email{jesper.christiansen@thep.lu.se}
\date{July 8, 2015}

\begin{abstract}
We present a series of observables for soft inclusive physics, and utilize them for comparison between two recently developed colour
  reconnection models; the new colour reconnection model in \pythia and the
  \dipsy rope hadronization model. The observables are ratios of identified
  hadron yields as a function of the final-state activity, as measured by the
  charged multiplicity. Since both considered models have a nontrivial dependence 
  on the final-state activity, the above observables serve as 
  excellent probes to test the effect of these models. Both models show a clear baryon
  enhancement with increasing multiplicity, while only the \dipsy rope model
  leads to a strangeness enhancement. 
  Flow-like patterns, previously found to be connected to colour reconnection
  models, are investigated for the new
  models. Only \pythia shows a $p_\perp$-dependent enhancement of the $\Lambda
  / K$ ratio as the final-state activity increases, with the enhancement being 
  largest in the mid-$p_\perp$ region.   
\end{abstract}

\preprint{LU-TP-15-26}
\preprint{MCNET-15-16}

\maketitle
%\flushbottom
%\sloppy

\section{Introduction}
The first run of the LHC has provided a large number of measurements probing both
soft and hard QCD, and thereby a large number of tests for the Monte Carlo
 event generators. Even
though the overall performance of the event generators have been quite good, there are
still some phenomena that are insufficiently understood~\cite{Roeck:2013jaa}. 
One of the more intriguing soft QCD deviations is the observed
enhancement of $\Lambda$
production~\cite{Aamodt:2011zza,Khachatryan:2011tm}. No model has been
simultaneously able to describe the identified hadron spectra at both LEP and
LHC. This has led to the development of several phenomenological
models~\cite{Bierlich:2014xba,Christiansen:2015yqa,Pierog:2013ria}, partly aimed
to  address this problem. With the planned low pile-up runs at the beginning
of the second LHC run, it is now an ideal time to test these models further, and thereby probe
the physical origin of the $\Lambda$ enhancement. In this study we 
consider two of the models: the new colour reconnection (CR) model in the
\pythia event generator \cite{Christiansen:2015yqa,Sjostrand:2014zea} and the
colour rope model in the \dipsy event generator
\cite{Bierlich:2014xba,Avsar:2005iz,Flensburg:2011kk}. The models have
previously been compared to $pp$ data at $\sqrt{s}$ of 200, 900 and 7000
GeV. In this paper new possible observables to test the models are suggested,
and predictions are made for collisions at $\sqrt{s} =$ 13 TeV. The observables are not model dependent, and can be used for constraining predictions from other models of soft inclusive physics. Both considered colour
reconnection models are built upon the Lund model for string
hadronization~\cite{Andersson:1983ia}. Nonperturbative differences can therefore be ascribed
to differences in the new phenomenological ideas.  

One of the key ideas for the two models in question is \textit{jet
  universality}. Stated in terms of the string model, it essentially means
that fragmentation of a string does not depend on how the string is 
formed. Free strings at both lepton and hadron colliders should
thus hadronize in a similar fashion. Fragmentation parameters
are therefore tuned in the clean $e^+e^-\rightarrow Z \rightarrow q\overline{q}$
environment, and then directly applied to hadron colliders. Any discrepancy
has to be due to physical phenomena not active at lepton colliders.
For all the models attempting to describe the $\Lambda$ enhancement, the
enhancement is linked to the increased density of quarks and gluons in the
final state at hadron colliders\footnote{Sometimes also referred to as string
  density, colour density, or energy density.}. It would therefore be of natural interest
to measure the $\Lambda$ enhancement as a function of this density. The
quark-gluon density is experimentally ill-defined, however, and we suggest to
use the number of 
charged tracks in the forward region as a measure of final-state activity. A
similar idea for using the hyperon-to-meson ratio to search for indications of
a miniQGP was suggested in ref.~\cite{Pop:2012ug}. We suggest ratios
that allows for separation of strangeness enhancement from baryon enhancement,
which both could be present in the hyperon-to-meson ratio.  

Another puzzling observation is the indication of collective effects in high-multiplicity $pp$ 
collisions~\cite{Kisiel:2010xy,Khachatryan:2010gv}, often interpreted as the presence of
flow. These effects were only expected in the dense medium of heavy ion collisions, where the pressure
gradients give rise to flow effects. A study of the models for $pp$ collisions
showed that CR generated similar effects even without the introduction of a
thermalized medium~\cite{Ortiz:2013yxa}. We therefore consider one of the standard
observables in heavy ion physics, that of identified particle ratios as a function of $p_\perp$, separated into
bins of centrality, and compare the model predictions for $pp$ collisions. Since centrality is not experimentally well defined in $pp$ collisions, the
number of charged tracks in the forward region is used as a measure of activity.

The outline of the paper is as follows. In section \ref{sec:models} we will
briefly recap the most important features 
of the two models considered. Comparison to existing $e^+e^-$ data at $\sqrt{s} = 91.2$ GeV, and $pp$ data at $\sqrt{s}$ of 200 GeV and 7 TeV, is shown in section \ref{sec:datacomp}. The event selection and tuning for 13 TeV is described
in section \ref{sec:tuning}. In section \ref{sec:predictions}, the predictions at $\sqrt{s}=$ 13 TeV, for
the second run of LHC, are presented. Finally, in section \ref{sec:conclusion},
we summarize and present an outlook.
 
\section{The models \label{sec:models}}
Both models for colour reconnection are built upon the Lund string model for
hadronization. In this model, outgoing partons are connected with stringlike
colour fields, which fragment into hadrons when moving apart.
The model contains two main parameters relevant to this study, which determine the suppression of strange
quarks and of diquarks (giving baryons) in the break ups. Assuming jet
universality, these parameters are tuned to LEP data. 

\begin{figure*}
\includegraphics[width=0.8\textwidth]{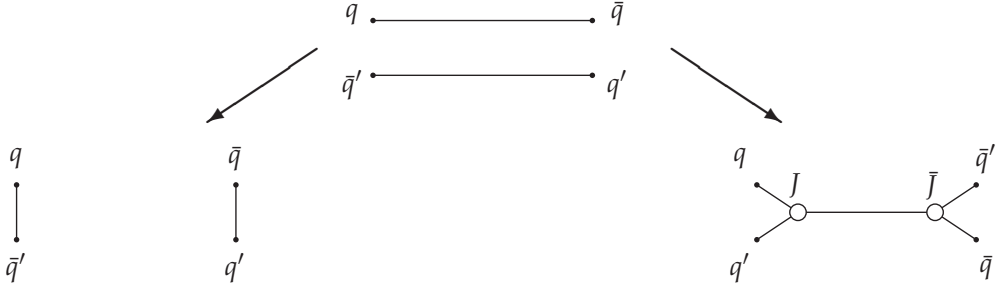}
  \caption{ \label{fig:colrec} Sketch of how two $q\bar{q}$ dipoles (top) can be
    reconnected to different colour topologies (left and right). The right
    connection gives rise to a double 
    junction, which in turn will produce baryons. Notice that the placement of
    the pairs differs in the junction figure.} 
\end{figure*}

Baryons can in addition be created around string junctions, which can arise as a
consequence of colour reconnection. Consider the simple configuration of two
$q\bar{q}$ dipoles in \figref{fig:colrec}, which for example could have
originated from a decay of two $W$-bosons in a LEP environment, as described in
ref.~\cite{Sjostrand:1993hi}. What essentially could be described as a
quadrupole configuration is instead described as either the original (on top) or the
left configuration in \figref{fig:colrec}. Without CR only the original
configuration is considered. Extending this type of colour
reconnection to hadron colliders has been shown \cite{Sjostrand:1987su} to be
a necessary condition to describe the rising of $\langle p_\perp
\rangle(N_{ch})$ distributions. The QCD $\varepsilon$-tensor gives rise to the
rightmost configuration, containing two junction connections, depicted as
empty circles. Since such junctions constitute proto-baryons, in the same way
string segments constitute proto-mesons, they become an additional source of
baryons. 

\subsection{Colour reconnection in \pythia}
The new CR model in \pythia is situated just prior to the
hadronization. It takes the leading-colour ($N_c \rightarrow \infty$) strings 
and transform them to a different colour configuration based on three
principles: firstly the 
SU(3) colour rules from QCD determine if two strings are colour compatible
(\eg there is only a $1/9$ probability that the top configuration of
\figref{fig:colrec} can transform to the left configuration purely from
colour considerations). Secondly a
simplistic space-time picture to check causal contact between the
strings. Finally the $\lambda$ measure~\cite{Andersson:1998tv} (which is a string-length measure, $\lambda = \sum_i \log(1+m_i^2/(2m_0^2))$ where the sum goes over all dipoles, $m_i$ is the invariant mass of the dipole and $m_0$ is a parameter) to
decide 
whether a possible reconnection is 
actually favoured. Since the model relies purely on the outgoing partons, it is in 
principle applicable to any type of collision. So far it has only
been tested for $pp$~\cite{Christiansen:2015yqa} and $ee$ collisions~\cite{Christiansen:2015yca}. The main extension
compared to the other CR models in \pythia is the introduction of
reconnections that form junction structures. From a pure colour consideration the
probability to form a junction topology is three times larger than an
ordinary reconnection. The
junction will introduce additional strings, however, and it is therefore often
disfavoured due to a larger $\lambda$ measure. Given the close connection
between junctions and baryons, the new model predicts a baryon enhancement.
It was shown to be able to simultaneously describe the $\Lambda$
production for both LEP and LHC experiments, which neither of the earlier
\pythia tunes have been able to. 

The new CR model essentially contains two new parameters: a parameter that
constrains the overall strength of the CR, and a parameter that controls the
baryon enhancement. Both of these parameters were tuned to data \cite{Khachatryan:2011tm,Aad:2010ac} from the LHC
experiments at 7 TeV.

\subsection{Rope hadronization in \dipsy}
Rope hadronization \cite{Biro:1984cf} is a normer for QCD inspired models, which includes interactions between strings. From previous attempts to include this effect in Monte Carlo generators \cite{Sorge:1992ej}, it is well known that strange and baryonic content will rise in very dense events.\par
A model introducing rope hadronization was recently developed and implemented in the event generator \dipsy \cite{Bierlich:2014xba}. Along with a final-state swing, the model introduces local calculation of string density, and corrects the evolution of the final-state parton shower and hadronization based on this local density.\par 
The model is based upon the idea \cite{Biro:1984cf} that when several parton pairs are next to
each other in geometric space, they can act together coherently to form a
colour rope. Each string is treated as a flux tube with a fixed radius, and
the amount of overlap between strings, in impact parameter space and rapidity,
can be directly calculated. \par 
If such an overlap is found to exist, it can have different effects, determined by SU(3) colour rules.
The overlapping strings can end up in a colour singlet configuration. This is handled by a final-state
"swing", that reconnects colour dipoles, in the final-state parton shower as
the transformation from the top to the bottom left configuration in \figref{fig:colrec}. 
In all other cases, the strings end up forming a "rope". This is hadronized with a higher effective string tension,
reflecting the fact that more energy is available for the fragmentation, in
accordance with results from lattice QCD \cite{Bali:2000un}. In some cases, the strings forming the rope end up in a junction structure. 
In such cases the junction pair is handled using a simple approach, where the
two junctions collapse to either two diquarks or two quarks, with a probability
controlled by a tuneable parameter. The resulting strings are then hadronized with the appropriate
effective string tension.\par 
An increased string tension results in more strange quarks and diquarks
produced in string breakups. Since the effect increases with the density of
quarks and gluons in the final state, the expected outcome is more baryons and
strangeness among the resulting hadrons. The model includes two free
parameters; the string radius and the probability for a junction to resolve to
diquarks. Both are tuned to LHC data \cite{Khachatryan:2011tm} at 7 TeV. 

\section{Comparison to data \label{sec:datacomp}}
\begin{figure}
	\centering
	\includegraphics[width=0.6\textwidth]{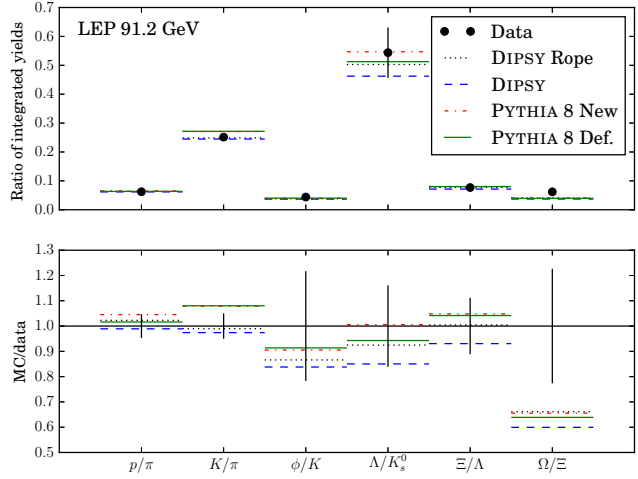}
	\caption{\label{fig:lepfig} Comparison to $e^+e^-$ at 91.2 GeV from ALEPH, SLD and PDG. Color online.}
\end{figure}

\begin{figure}
\centering
\includegraphics[width=0.49\textwidth]{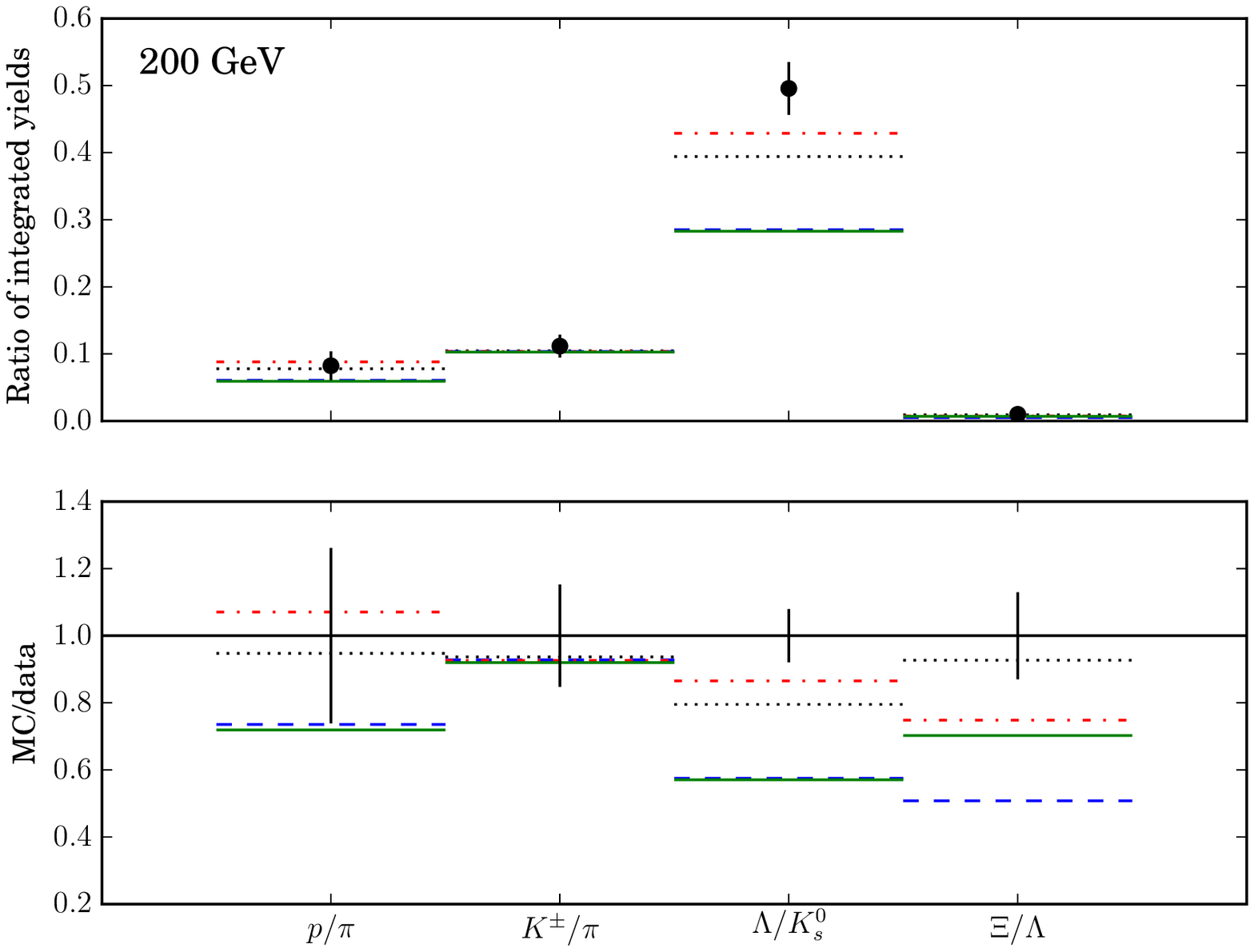}
\includegraphics[width=0.49\textwidth]{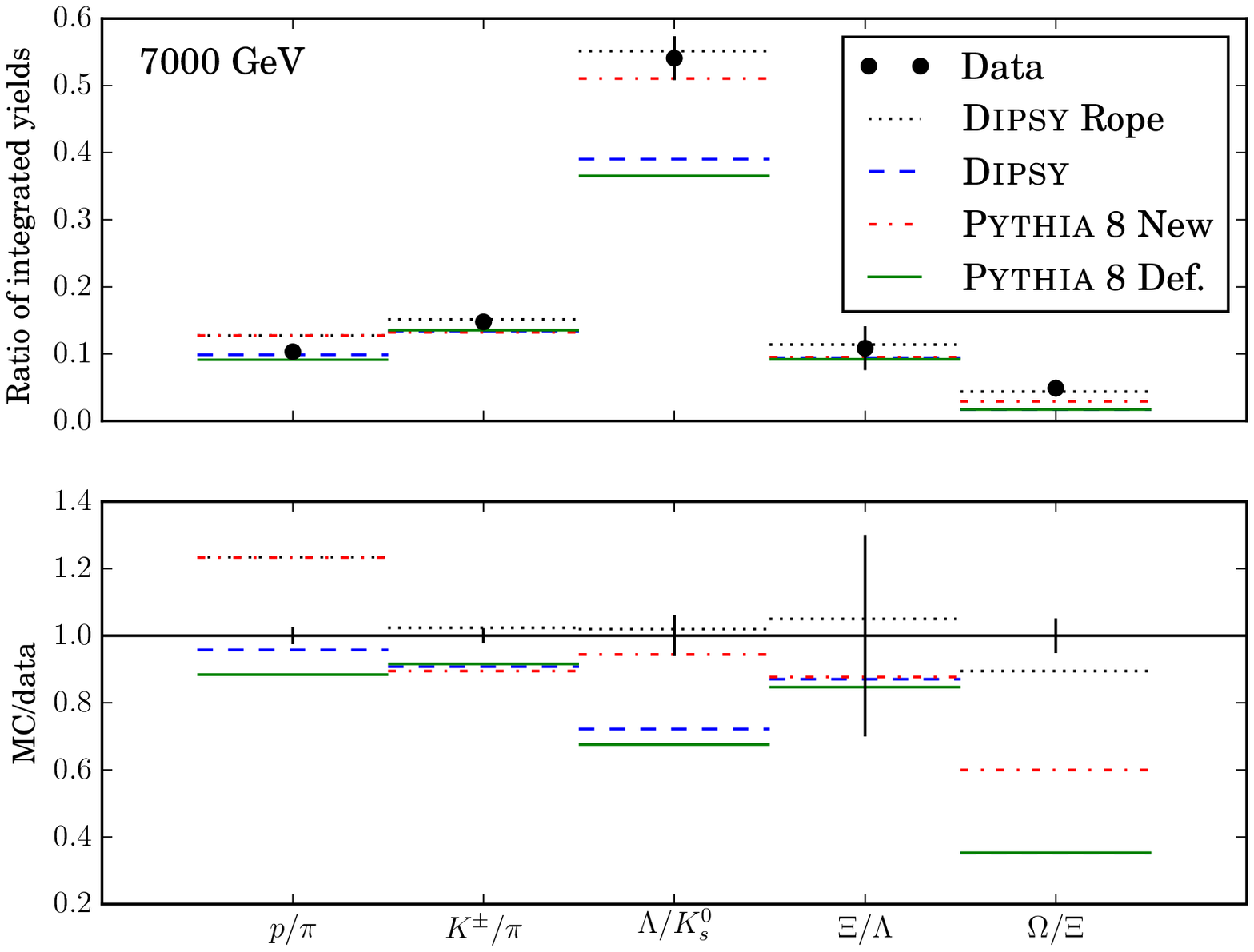}
\caption{\label{fig:datacomp} Comparison to $pp$ data at 200 GeV from STAR (left) and at 7 TeV from ALICE and CMS (right). Color online.}
\end{figure}

The models performs as intended when comparing to existing data. Ratios of
baryons to mesons are enhanced for both models, whereas ratios of particles
with strange quark content is enhanced only in the \dipsy rope
model. Comparisons are done to ratios of integrated yields of identified
particles, using the analyses published through the
Rivet~\cite{Buckley:2010ar} framework. The raw results from comparing the
Monte 
Carlo to data using Rivet, are integrated to give figures~\ref{fig:lepfig} and~\ref{fig:datacomp},
using Matplotlib~\cite{Hunter:2007}. Error estimates are conservative, as they
assume the error of all bins are fully correlated.\par 
In figure \ref{fig:lepfig}, a comparison to LEP data \cite{Barate:1996fi,
  Abe:2003iy, Amsler:2008zzb} is seen. Two conclusions can be drawn from this
figure. First of all, these are the data the original string model is tuned
to. The fact that the Monte Carlo is so well aligned with data is thus not an
indication that the string model predicts all these ratios so well, but rather
that the parameters of the model are tuned to these data. The exception here is
the $\Omega$ baryon\footnote{We denote a particle and its antiparticle with
  just a single letter such that \eg $p$ means both proton and
  anti-proton. Special cases are $\pi$ with denotes $\pi^+\pi^-$, $K$ which
  denotes 
  $K^+K^-K^0_sK^0_L$ and $\Xi$ which denotes $\Xi^+\Xi^-$.}, reflected in the
$\Omega/\Xi$ ratio, which lies below the observed value. However, the
experimental statistical uncertainty is large for this ratio.\par 
The other conclusion, which is the most relevant for this article, is that
only small effects at LEP data is observed. The $\Lambda / K$ ratio
increases by about 10~\% for both the DIPSY rope model and the new CR model in
\pythia (over their respective default models), but all models stay within the experimental
uncertainty. The overall low variance is exactly what is expected, due to the
low final state activity at LEP.

In figure~\ref{fig:datacomp} comparison to STAR
data~\cite{Abelev:2008ab,Abelev:2006cs,Adams:2006nd} at 200 GeV indicated that
the description of the baryon to meson ratios  improves with both models, while the
description of the $\Xi/\Lambda$ ratio only improves with the \dipsy rope
model. The change in the $K^\pm/\pi$ ratio is not visible on this scale for this
energy.\par 
Comparison to 7 TeV data from ALICE \cite{Adam:2015qaa,Abelev:2012jp} and
CMS~\cite{Khachatryan:2011tm} confirms that the 
description improves, even for the $\Omega/\Xi$ ratio. The description of the
$p/\pi$ ratio is seen to be somewhat worse with the new models. This could
either have a mundane explanation originating in the fact that the very
low-$p_\perp$ area of the individual distributions (which contains most of
the multiplicity) are not fully understood, or have further reaching
consequences. We point to the measurements suggested in the next section of
this paper to shed light on this issue. 

\section{Tuning and event selection \label{sec:tuning}}
Before studying exclusive observables at 13 TeV, it is necessary to verify that the 
baselines for the two models agree reasonable well. Normally
this is achieved by tuning the models to the available data. Data at $\sqrt{s} = 13$ TeV is, however, not yet published in a state where event generators can be tuned to it, so the \dipsy model was instead tuned to the \pythia
predictions for $\text{d}N_{ch}/\text{d}\eta$, $\langle p_\perp \rangle \,
(N_{ch})$ and the multiplicity distribution. Both models will eventually have
to be retuned, when more data, in a suitable format for tuning, become
available. Only small effects are expected from the retuning,  
firstly due to fragmentation mainly being determined from LEP data, and
secondly since the already presented results at 13 TeV show a good agreement
between the Monash tune and the data~\cite{1384979,ATLASue}. The full list of
all parameters changed from their default values is included in an appendix.

An event and particle selection was implemented to mimic a possible experimental
setup. Each particle is required to have $p_\perp > 0.15~\text{GeV}$. Two
different $\eta$ regions are used; a forward region ($2 < |\eta| < 5$)
to measure the activity, and a central region  ($|\eta| < 1$) to measure
the identified hadron yields. The reason for the split is to avoid any potential
bias, which otherwise happens at low $N_{ch}$, in particular for ratios involving
both charged and non-charged hadrons. Since \dipsy does not have
a model for diffraction, only non-diffractive events are considered for both
models. To reflect this in the event selection, only events with at least six forward
charged particles are considered.

All particles with $c\tau > 10$ mm are treated as stable. In practice this
means that $\pi^\pm, K, \Lambda,
\Xi$ and $\Omega$ are all stable whereas $\phi$ (which decays strongly) is
not. This introduces some double counting in the $\phi/K$-ratio, where a
$\phi$ can potentially be counted in the numerator and its decay products in
the denominator. 

\section{Predictions for 13 TeV \label{sec:predictions}}
Differences between the colour reconnection models are best determined using
observables controlled by hadronization effects. Ratios of identified
particles is exactly such an observable, since particle species production is
 determined by the quark and diquark content in string breaks. In the
first part of this section, ratios of identified particles are shown as a
function of $N_{ch}$ in the forward region, as a measure of event
activity. Then flow-like effects are considered, by showing $(\Lambda /
K)\,(p_\perp)$ in four different bins of $N_{ch}$ in the forward region. 

\begin{figure*}
\centering
\includegraphics[width=\textwidth]{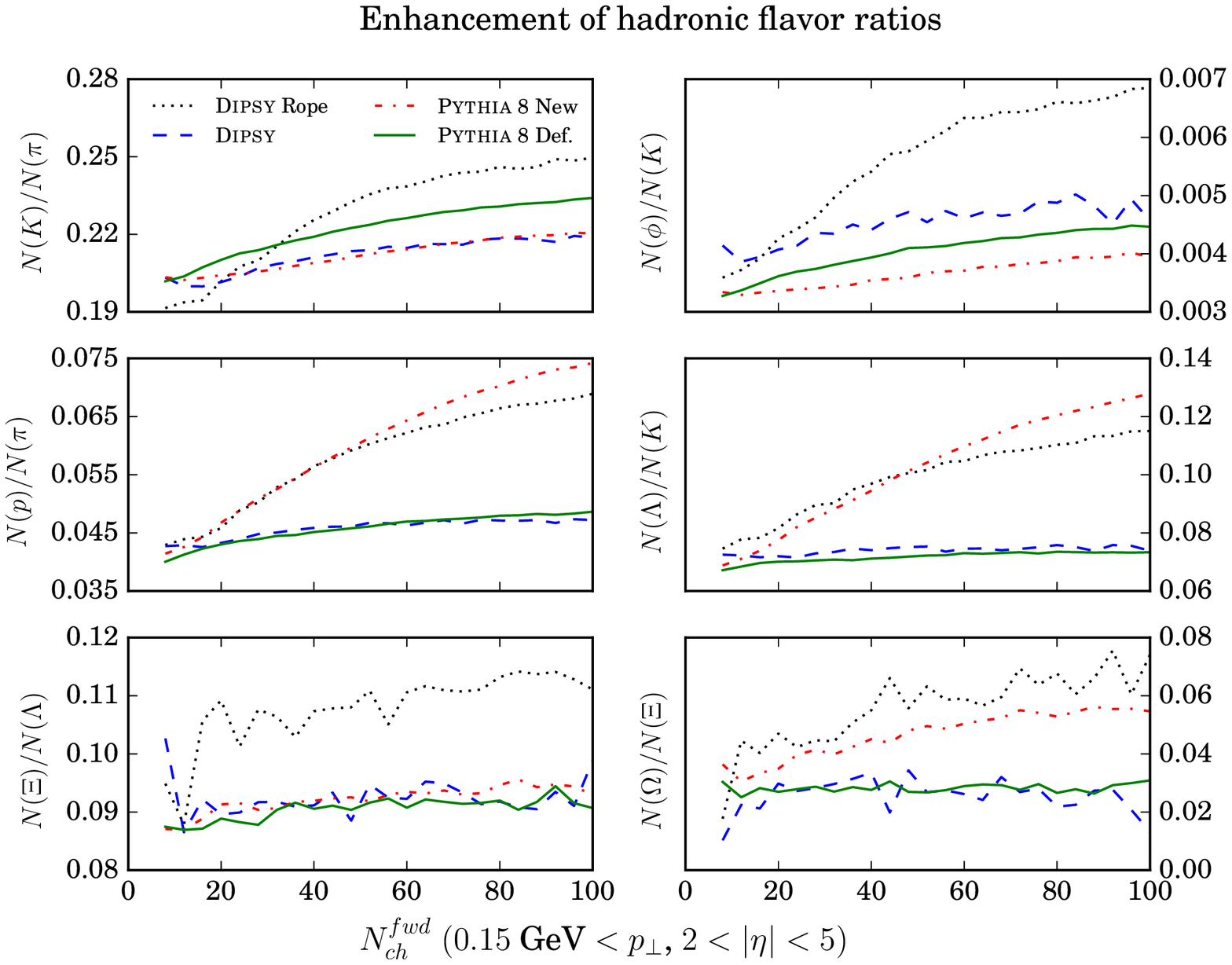}
\caption{Ratios of identified hadrons as functions of $N_{ch}^{fwd}$ at
  $\sqrt{s} = 13$ TeV. The top row shows meson ratios with the numerator
  having one more strange quark than the denominator. The middle row shows baryon to meson
  ratios, with same amount of strange quarks. The bottom row shows baryon
  ratios with the numerator
  having one more strange quark than the denominator. Note that the vertical
  axis differs between the figures and that zero is suppressed.\label{fig:hadronratios}} 
\end{figure*}

\subsection{Particle ratios \label{sec:ratios_nch}}
Ratios of hadrons with different strange and baryon numbers as function of
event activity, measured as functions of $N_{ch}^{fwd}$, are shown in
\figref{fig:hadronratios}. The strangeness enhancement in meson production is
probed by the $K/\pi$ and $\phi/K$ ratios, for which the numerator always
contains one more strange quark than the denominator. As expected, only the
\dipsy rope model shows an enhancement relative to the baseline, since it
contains a strangeness enhancement. The new \pythia CR model lies slightly
below the baseline. This can be explained by phase-space constraints for low
invariant-mass strings, which the new model produces more of. It should be
recalled that both the new as well as the old models are 
capable of describing the total $K_s^0$ yield at 7 TeV. Thus, the limited
effects in this ratio is somewhat expected. The $\phi/K$ ratio shows more
promise as a means to distinguish between the two models, since the \dipsy
model shows a larger enhancement. It is, however, more experimentally
challenging.

The baryon enhancement is tested for both hadrons containing zero or one
strange quark, $p/\pi$ and $\Lambda/K$. For both ratios, and both models, clear
enhancements are expected and seen. For the $\Lambda/K$ ratio both models
agree quite well, which is not surprising, given that both models are tuned
to describe the inclusive $\Lambda/K$ distributions at 7~TeV. A similar
picture is seen for the $p/\pi$ ratio, indicating similar predictions for the
baryon enhancement from both the models.

The multistrange baryon enhancement is tested in the same way as the strange-meson
enhancement by considering the ratios $\Xi / \Lambda$ and $\Omega /
\Xi$. The large variations at low multiplicity for both distributions are
due to statistical fluctuations. For $\Xi / \Lambda$ the \dipsy rope model shows a clear enhancement as opposed to the new \pythia CR
model. The $\Lambda / p$ ratio is not shown, but the enhancement is similar to the enhancement of $\Xi / \Lambda$.
An enhancement is seen for both models in the $\Omega / \Xi$, with the enhancement
factor being around 2.5 for the \dipsy rope model in the highest multiplicity
bins. This is larger than any of the other enhancements
seen. The enhancement for the new \pythia CR model is somewhat surprising, as
the increased junction production should be equal for both $\Xi$ and
$\Omega$. The production of $\Omega$ in the standard \pythia fragmentation is,
however, significantly suppressed, as the production of $ss$-diquarks is
disfavoured. This suppression is not present in the junction handling, since
it takes two already formed quarks and combine into a diquark. The enhancement in the new \pythia
model should therefore not be interpreted as a "real" strangeness enhancement,
but more as an absence of suppression of $ss$ diquarks. For the \dipsy model
the above effect is also present, but there is an additional enhancement
of strangeness and diquarks. It should be noted that the $\Omega$ baseline
from LEP is not that well constrained, due to a large experimental
uncertainty, and the model predictions are below the actual 
measurements. A measurement of $(\Omega/\Xi) \,(N_{ch})$
would cast light on whether an actual activity-based enhancement takes
place. 

Increased hyperon production in high activity $pp$ events have
previously been associated with production of a miniQGP \cite{Pop:2012ug}. The
hyperon-to-pion ratio is only indirectly shown in \figref{fig:hadronratios},
but the rise is similar to the one predicted by miniQGP. The new models
therefore provide an alternative explanation, if such an enhancement is
observed. 

\subsection{Flow-like effects}
\begin{figure*}
\centering
\includegraphics[width=0.9\textwidth]{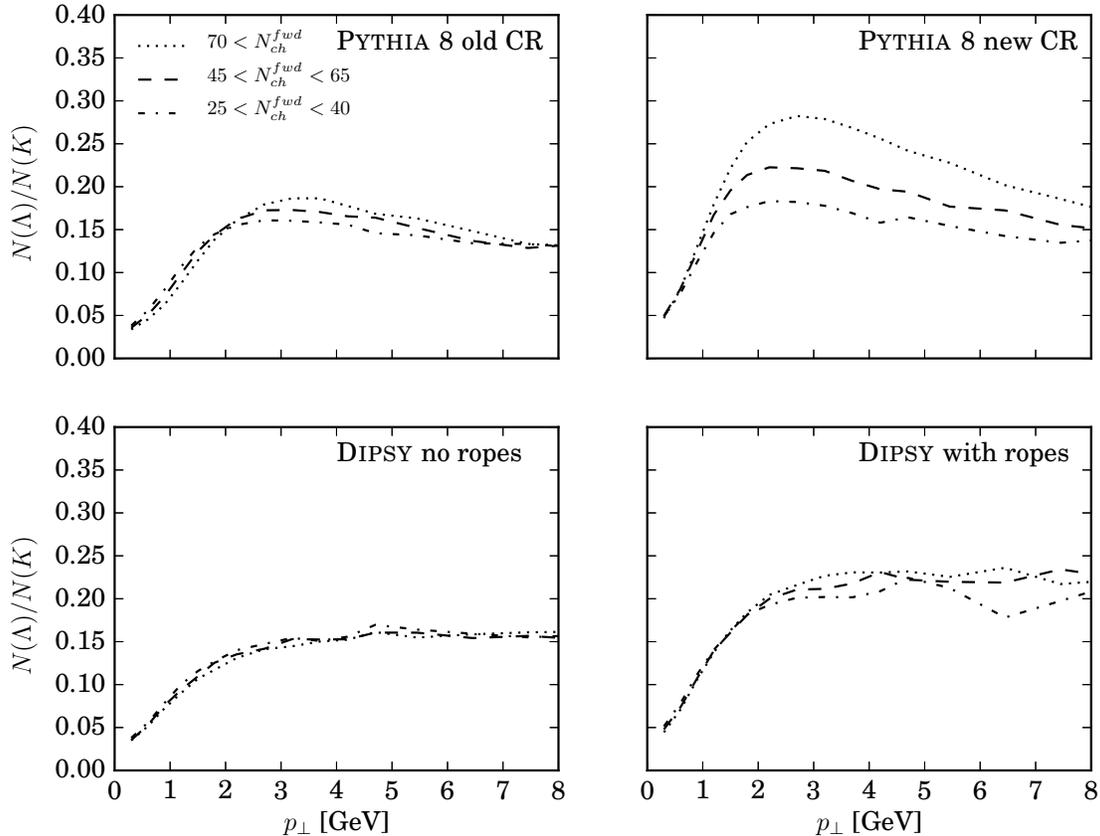}
\caption{\label{fig:flowlike} Ratio of $\Lambda/K$ as a function of $p_\perp$ in three bins of $N_{ch}^{fwd}$. In the right column the new colour reconnection models are shown, and in the left column the old ones.}
\end{figure*}

The $\Lambda / K$ ratio as a function of $p_\perp$ for different
$N_{ch}^{fwd}$ ranges is shown in \figref{fig:flowlike}. The two models show
different behaviours for the different multiplicity ranges: 
the \dipsy rope model only gives a small enhancement ($\sim 10 \%$ at maximum)
between the lowest and highest multiplicity regions. Even though the
differential enhancement is generally below 10 \%, the enhancement of the
ratio of integrated yields is about 20 \%, which is in good agreement with
\figref{fig:hadronratios}. It should 
be noted that the \dipsy model is 
inadequate in describing the high $p_\perp$ tails ($p_\perp >\sim 4$
GeV). This was observed for 900 GeV and 7 TeV in ref.~\cite{Bierlich:2014xba}.

The new \pythia CR model shows a clear change in $p_\perp$ with increasing
multiplicity. The enhancement is largest in the mid-$p_\perp$ region
($p_\perp \sim 2-6$ GeV),
leading to a "peak" structure. This structure looks qualitatively
similar to what is observed in $PbPb$ and $pPb$ collisions~\cite{Abelev:2013xaa,Abelev:2013haa}. The
peak also moves towards larger $p_\perp$ with increased multiplicity, an effect
normally attributed to radial flow in heavy ion
collisions~\cite{Fries:2003fr}. That the new CR model predicts a qualitatively
similar effect in $pp$ collisions is quite intriguing and strengthens the hint at a
potential connection between flow and CR effects already
observed~\cite{Ortiz:2013yxa}. 

\section{Conclusions\label{sec:conclusion}}
A series of new model-independent observables, well suited for distinguishing
between between different physical models for soft inclusive physics is
suggested. The observables are ratios of identified hadrons measured as a
function of event activity, with the identified hadrons chosen such that a
distinction is made between baryon-only, strangeness-only and
baryon-and-strangeness enhancement. Measurement of these observables at
present and future energies at $pp$ colliders is encouraged, as the can serve
as constraints on any soft physics model aiming to explain low-multiplicity
and minimum bias data simultaneously.\par 
The observables are, in this article, used to separate two new CR models. 
The new CR model in \pythia only contains a baryon enhancement
with increasing multiplicity, while the \dipsy rope models contains both a
baryon and a strangeness enhancement. The multistrange hyperon ratios, as
well as the $\phi/K$ ratio, provide clear observables for distinguishing
between the two models. It should be mentioned that this is already possible to
observe in inclusive measurements, but the separation into different
multiplicity regions highlights the enhancement.

Both new models are based on interactions between strings in the hadronization
phase, and confirmation of the common predictions made by the two models is a
direct hint that colour reconnections among strings are of physical
importance. 
Both baseline models show almost no dependency on multiplicity for the
identified hadron yield ratios. Therefore, any observed dependency would
provide a clearer indication that the old models miss a feature, better than
an inclusive measurement alone could provide. We therefore strongly suggest that
these observables should be measured at the LHC experiments. In this paper we only
studied the effects at a center-of-mass energy of 13 TeV, but the effects
should also be visible in the already collected data at 7
TeV. 

We have also shown that one of the CR models predicts effects similar to those
normally attributed to radial flow in heavy ion collisions. This is in agreement
with earlier indications that also hint at a connection between the two
phenomena. It should however be recalled that neither of the models provide a
satisfactory description of the individual $p_\perp$ spectra for the
identified hadrons. And before these are fully understood, claims of connections between flow and CR may be premature.

\section{Acknowledgments}
We thank Leif L{\"o}nnblad and Torbj{\"o}rn Sj{\"o}strand for useful
discussions and comments.
Work supported in part by the MCnetITN FP7 Marie Curie Initial
Training Network, contract PITN-GA-2012-315877 and the Swedish Research
Council, contract 621-2013-4287.

\appendix
\section{Model parameters}
A complete list of all the parameters that differ from their default values
for the considered models. \\
Hadronization model parameters are found in table \ref{tab:hadtable}, \textsc{Pythia} parameters in table \ref{tab:pytpar} and \dipsy parameters in tables \ref{tab:ropepar} and \ref{tab:dippar2}.
\begin{table}
\begin{minipage}{\linewidth}
\centering
  {\ttfamily\selectfont
   \begin{tabular}{lcccr}
   \toprule
   \textnormal{Fragmentation parameter} & \textnormal{\textsc{Pythia} def.} & \textnormal{\textsc{Pythia} new} & \textnormal{\dipsy} & \textnormal{\dipsy rope} \\
    \hline
    StringPT:sigma                        &  0.335 &  0.335 &  0.32 &  0.31 \\
    StringZ:aLund                         &  0.68  &  0.36  &  0.30  &  0.41  \\
    StringZ:bLund                         &  0.98  &  0.56  &  0.36  &  0.37 \\
    StringFlav:probQQtoQ                  &  0.081 &  0.078 &  0.082 &  0.073   \\
    StringFlav:ProbStoUD                  &  0.217 &  0.22  &  0.22   &  0.21 \\
    \multirow{4}{*}{StringFlav:probQQ1toQQ0join} &  0.5 &  0.0275&-&- \\
    &  0.7 &  0.0275&-&- \\
    &  0.9 &  0.0275&-&- \\
    &  1.0 &  0.0275&-&- \\
    \botrule
    \end{tabular}
    }
    \end{minipage}
    \caption{\label{tab:hadtable} Table of parameters of the string hadronization model, which differs from Monash tune \cite{1384979} default values. The changed parameters have been retuned to LEP and SLD data, cf. figure \ref{fig:lepfig}}.
    \end{table}
    \begin{table}
    \begin{minipage}{\linewidth}
\centering
  {\ttfamily\selectfont
   \begin{tabular}{lcr}
   \toprule
   \textnormal{\textsc{Pythia} parameter} & \textnormal{Default} & \textnormal{New} \\
    \hline
    MultiPartonInteractions:pT0Ref        &  2.28  &  2.15 \\ 
    \hline
    BeamRemnants:remnantMode              &  0     &  1  \\
    BeamRemnants:saturation               & -       &  5   \\
    \hline 
    ColourReconnection:mode               &  0     &  1  \\
    ColourReconnection:allowDoubleJunRem  &  on    &  off \\
    ColourReconnection:m0                 & -       &  0.3 \\
    ColourReconnection:allowJunctions     & -       &  on  \\
    ColourReconnection:junctionCorrection & -       &  1.2 \\
    ColourReconnection:timeDilationMode   & -       &  2    \\
    ColourReconnection:timeDilationPar    & -       &  0.18 \\
    \botrule
   \end{tabular}
    }
    \end{minipage}
    \caption{\label{tab:pytpar} The new \textsc{Pythia} CR model introduces a number of new parameters, and requires retuning of a few old ones, besides hadronization. The details of the retuning can be found in ref. \cite{Christiansen:2015yqa}. }
    \end{table}
\begin{table}   
\begin{minipage}{\linewidth}
\centering
  {\ttfamily\selectfont
   \begin{tabular}{lcr}
   \toprule
   \textnormal{\dipsy parameter} & \textnormal{Default} & \textnormal{Rope} \\
   \hline
   FragmentationScheme   & default & dipole \\
    StringR0              & -     &  0.773  \\
    Stringm0              & -     &  0.113   \\
    BetaPopcorn	    & -     &  0.2 \\
    \botrule
   \end{tabular}
    }
    \end{minipage}
    \caption{\label{tab:ropepar} The \dipsy rope model introduces three extra parameters, which are fixed using $pp$ data from LHC. See ref. \cite{Bierlich:2014xba} for the meaning of the parameters.}
    \end{table}
    \begin{table}
\begin{minipage}{\linewidth}
\centering
  {\ttfamily\selectfont
   \begin{tabular}{lccccccr}
   \toprule
   \textnormal{Energy} & \multicolumn{2}{c}{$pp$\textnormal{ 200 GeV}} &  \multicolumn{2}{c}{$pp$\textnormal{ 7 TeV}} & \multicolumn{2}{c}{$pp$\textnormal{ 13 TeV}}  \\
   \hline
   \textnormal{\dipsy parameter} &  \textnormal{Default} & \textnormal{Rope}&  \textnormal{Default} & \textnormal{Rope}&  \textnormal{Default} & \textnormal{Rope}\\
    \hline
    LambdaQCD & 0.29 & 0.26 & 0.17 & 0.25 & 0.29 & 0.27 \\
    RMax & 2.32 & 3.34 & 3.23 & 2.90 & 1.05 & 3.39 \\
    PMinusOrdering & 1.05 & 0.98 & 1.24 & 0.67 & 0.31 & 0.75 \\
    PTScale & 0.70 & 0.92 & 1.60 & 1.65 & 1.28 & 1.35 \\
    \botrule
  \end{tabular}
 }
\end{minipage}
\caption{\label{tab:dippar2} The \dipsy initial state model needs retuning at each energy to reproduce total charged multiplicity. See ref. \cite{Flensburg:2011kk} for  the meaning of the parameters.}
\end{table}
\clearpage
\bibliography{bibliography}
\end{document}